\documentclass[twocolumn,showpacs,preprintnumbers,amsmath,amssymb,prb]{revtex4-1}

\usepackage[active]{srcltx}
\usepackage{graphicx}
\usepackage{dcolumn}
\usepackage{bm}
\usepackage{mathtools}

\usepackage{hyperref}
\usepackage{color}
\usepackage[a4paper]{anysize}
 
\renewcommand{\a}{{\bf a}}
\renewcommand{\k}{{\bm{k}}}
\newcommand{\E}{{\bf E}}
\renewcommand{\b}{{\bf b}}

\begin{document}

\title{Tight-binding approach to penta-graphene}
\author{T. Stauber,$^1$ J. I. Beltr\'an,$^{1,2}$ and J. Schliemann$^3$}
\affiliation{
$^1$Instituto de Ciencia de Materiales de Madrid, CSIC, 28049 Madrid, Spain\\
$^2$GFMC and Instituto Pluridisciplinar, Departamento de F\'isica Aplicada III, Universidad Complutense de Madrid, 28040 Madrid, Spain.\\
$^3$Institute for Theoretical Physics, 
University of Regensburg, D-93040 Regensburg, Germany}

\begin{abstract}
We introduce an effective tight-binding model to discuss penta-graphene and present an analytical solution. This model only involves the $\pi$-orbitals of the sp$^2$-hybridized carbon atoms and reproduces the two highest valence bands. By introducing energy-dependent hopping elements, originating from the elimination of the sp$^3$-hybridized carbon atoms, also the two lowest conduction bands can be well approximated - but only after the inclusion of a Hubbard onsite interaction as well as of assisted hopping terms. The eigenfunctions can be approximated analytically for the effective model without energy-dependent hopping elements and the optical absorption is discussed. We find large isotropic absorption of up to 24\% for transitions at the $\Gamma$-point, but the general absorption will show a strongly anisotropic behaviour depending on the linear polarization of the incident light. 
\end{abstract}


\maketitle

\section{Introduction}
Arguably, carbon is the most versatile element being capable to form various stable structures with graphene\cite{Novoselov04,NovoselovPNAS05} being its most prominent two-dimensional allotrope out of which most carbon structures can be built such as fullerene,\cite{Kroto85} carbon-nanotubes\cite{Iijima91} or multi-layer graphene\cite{Neto09} and graphite.

Recently, a new allotrope was proposed which cannot be composed from a graphene sheet: penta-graphene.\cite{Zhang15} It entirely consists of carbon atoms forming pentagons within the Cairo patterning and it remains almost flat by adding to its sp$^2$-hybridised carbon atoms also sp$^3$-hybridised carbon atoms, arranging them in three parallel horizontal planes separated by approximately half an Angstrom. Penta-graphene, therefore, only experiences a buckling on the atomic scale of a unit cell and would represent another example of a two-dimensional semiconductor.

Even though penta-graphene has not been synthesised experimentally, the theoretical analysis point out many intriguing properties which are worth discussing in more detail. Most strikingly in terms of applications is probably the large predicted band-gap of 3.25eV which makes it a potential candidate for blue absorption/emission. Nevertheless, the optical properties such as absorption due to a linearly polarised light field have not been addressed, yet.

The objective of this work is two-fold. First, we will discuss the possibility of a simple tight-binding description to model the valence and conduction bands closest to the neutrality point. Our model will only include the $\pi$-orbitals of the sp$^2$-hybridised carbon atoms for which an analytical solution is possible. This is reminiscent to widely-used tight-binding models for graphene\cite{Wallace47,McClure56,Slonczewski58} and carbon-nanotubes.\cite{Jorio01,Reich02} Second, using the analytical approximation, we will also determine the absorption of linearly polarized light via Fermi's Golden Rule.

The paper is organised as follows. In Sec. II, we discuss the general structure of penta-gaphene. In Sec. III, we introduce the effective 4-band model and present its analytical approximation. In Sec. IV, we discuss the absorption at the high-symmetry points of the Brillouin zone and close with a summary and conclusions. In an appendix, we point out the importance of correlation effects in order to justify the parameters of the extended tight-binding model.

\section{Penta-graphene}
 
\subsection{Lattice structure}
The regular Cairo pentagonal patterning is characterized by four bonds forming the pentagon, three of which have the distance $a$ and one has the distance $b=(\sqrt{3}-1)a$. The unit cell consists of six carbon atoms and is defined by the two lattice vectors $\a_1=a(\sqrt{3},\sqrt{3})$ and $\a_2=a(\sqrt{3},-\sqrt{3})$. The length of the quadratic unit cell $|\a_1|=|\a_2|=\sqrt{6}a$ is obtained from first-principle studies to be $3.64$\AA,\cite{Zhang15} which translates into $a=1.49$\AA  and $b=1.09$\AA. 

The real distances as obtained from first-principle calculations turn out to be slightly different, i.e., $C1-C2=1.55$\AA  and $b=C2-C2=1.34$\AA, where there are two C1-atoms and four C2-atoms denoting the carbon atoms with sp$^3$ and sp$^2$-hybridization, respectively. Furthermore, the atoms are arranged in three different horizontal planes, of which the C1-atoms form the central plane and two of the four C2-atoms the upper and the lower plane, respectively.The total distance between the C1 and C2-atomic horizontal planes is $h=0.6$\AA which yields the projected 2D-distance $a=C1-C2 (projected)=1.43$\AA and $b=C2-C2=1.34$\AA. 

\begin{figure}
\centering
\includegraphics[width=0.99\columnwidth]{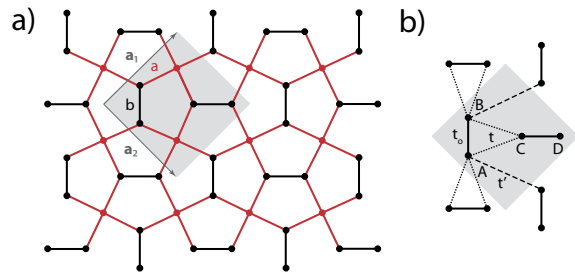}
\caption{(color online).  a) The lattice structure of penta-graphene resembling the Cairo pentagonal patterning. The black dots represent carbon atoms with sp$^2$-hybridization (C2) and the red dots stand for carbon atoms with sp$^3$-hybridization (C1). b) Unit cell of the reduced tight-binding model consisting of the four C2-atoms labeled as $A, B, C, D$ connected by the three hopping terms $t_0$, $t$, and $t'$.}
  \label{Cairo}
\end{figure}

The distorted Cairo pentagonal patterning is shown on the left hand side of Fig. \ref{Cairo} where the C1 and C2-atoms are represented by red and black dots, respectively. The black horizontal and vertical bonds of length $b$  connect the C2-atoms whereas the red bonds of the projected length $a$ connect the C1 with the C2-atoms.\footnote{There is a slight difference of 1\% between the two bonds connecting the C1-C2 atoms that form a pentagon. This will be neglected.We also assume that the projected distance $C1-C2(projected)$ is the distance $a$ of the regular Cairo patterning.} The unit cell is denoted by the shaded square and consists of two C1-atoms and four C2-atoms. 

Let us finally comment on the symmetry group. The three-dimensional lattice possesses the S$_4$ point group and $D_{2d}$ full space group. The latter includes the following symmetry elements: one C2 axis along the direction perpendicular to the $\a_1$-$\a_2$ plane, two C2' axes perpendicular to C2, two dihedral planes $\sigma_d$ bisecting the angles formed by pairs of the C2' axes and two improper S$_4$ axes. For the strictly two-dimensional lattice, the symmetry elements is doubled going from $D_{2d}$ to $D_{4h}$ full space group.

\subsection{Density Functional Theory calculations}
We calculate the band structure of pentagraphene using the VASP code,\cite{Kresse93,Kresse94} based on density functional theory (DFT). All calculations were done using the Projector Augmented Wave potentials\cite{Kresse99} and employing the Perdew-Burke-Ernzerhof flavour of the generalized gradient approximation for the exchange-correlation functional.\cite{Perdew96} Plane waves with a energy cut-off of 500 eV were employed to describe the valence electrons (2$s^2$ 2$p^2$) of the C atoms. The employed Brillouin zone for the relaxation calculations is 9x9x1 in the Monkhorst-Pack scheme,\cite{Monkhorst76} while high accurate electronic calculations to estimate the relative C1/C2 weight were performed using 11x11x1. The band structure of free standing pentagraphene layer was calculated for a fixed lattice constants of a=b=3.64\AA,\cite{Zhang15} and including a vacuum distance of around 20\AA. In all calculations the atomic positions are relaxed till forces are less than 0.015 e-/\AA.

We find slightly larger atomic distances than the one cited in the previous subsection which were taken from Ref. \onlinecite{Zhang15}.The obtained band structure shows the expected narrowing of the band gap, common to all local and semi-local exchange-correlation approximations, which is corrected via a rigid shift of the conduction band to mimic the obtained direct $\Gamma$-$\Gamma$ band gap using hybrid functionals.\cite{Zhang15} After the rigid shift, the band dispersion shown in Fig. 2  as red stars agrees remarkably well to the more costly hybrid functionals. The C1/C2 atomic contribution ratio of the valence and conduction band is less than 9\% and 10\%, respectively, so we can conclude that a four-band model is an adequate approximation.

\section{Tight-binding and analytical approach}
Our goal is to introduce a tight-binding model that only considers the four C2-atoms where the atoms $A$ and $B$ form the vertical dimer and the atoms $C$ and $D$ the horizontal dimers, see right hand side of Fig. \ref{Cairo}. Both dimers are coupled by the hopping matrix element $t_0$ which connects the two $\pi$-orbitals. We will set this to the typical value $t_0=2.7$eV.\cite{Neto09} The other hopping matrix elements involve hopping processes between the two dimers $t$ and next-nearest hopping processes between the same dimers $t'$. To simplify our model, we will set $t=t'$ and determine its value from a fit to the DFT-band structure. As we will show, this model can also be well described analytically. But first, we will discuss the full tight-binding model including the six atoms of the unit cell and 4 orbitals  within the Slater-Koster formalism.\cite{SlaterKoster54}

\subsection{Slater-Koster approach}
\begin{figure}[t]
\centering
  \includegraphics[width=1.05\columnwidth]{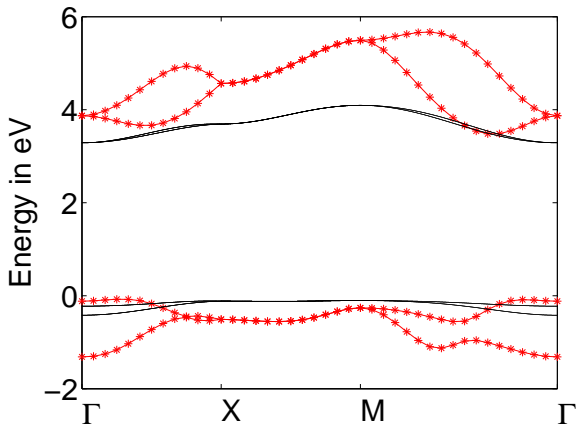}
\caption{(color online). The band structure obtained from the full Slater-Koster tight-binding model (black lines) compared to the DFT-band structure (red stars).}
  \label{CompareSlaterKoster}
\end{figure}

We start our tight-binding description with the full model that will include all six atoms of the unit cell and all four orbitals  of the second atomic energy level, i.e., $2s$, $2p_x$, $2p_y$, $2p_z$. The hopping parameters are chosen to be the one of graphene scaled by the corresponding distance and taken from Ref. \onlinecite{Yuan15}:  $V_{ss\sigma}=-5.34$eV, $V_{sp\sigma}=6.40$eV, $V_{pp\sigma}=7.65$eV, and $V_{pp\pi}=-2.80$eV. The atomic energy levels of the sp$^2$-hybridized C2-atoms are $\epsilon_s=-2.85$eV, $\epsilon_{p_{x/y}}=3.20$eV, $\epsilon_{p_z}=0$; the ones of the sp$^3$-hybridized C1-atoms are chosen as $\epsilon_s=-4.05$eV, $\epsilon_{p}=2$eV. The $24\times24$ Hamiltonian is obtained following the Slater-Koster parametrization including only nearest-neighbor hopping.\cite{SlaterKoster54}

In Fig. \ref{CompareSlaterKoster}, the lowest energy bands obtained from the full tight-binding model are shown (black lines) and compared to the DFT-results (red stars). Qualitatively, the tight-binding can reproduce the lowest conduction bands and the highest valence bands. Still, quantitatively there are several disagreements: i) the energy gap is not accurately reproduced; ii) the splitting of the two conduction as well as of the two valence bands is not small; iii) bands further away from the Fermi level become worse and they can at most only be described qualitatively. 

The above analysis does not contain any fitting parameter. Nevertheless, the quantitative disagreement suggests other terms to become important. In fact, in the case of the effective four-band model, we will be only able to adequately describe the bands by including an onsite interaction as well as an assisted hopping term. This shall be discussed in the following

\subsection{Effective four-band model}
As outlined above, we will build up an effective tight-binding model by only considering lattice sites with an unbounded $\pi$-electron. There are thus four atoms in the unit cell and direct hopping ($t_0$) within the unit cell is only between the dimers A-B and C-D, respectively. All other hopping processes involve intermediate lattice sites with sp$^3$-hybridisation and we assume that this hopping is the same ($t=t'$) and considerably less than the direct $\pi-\pi$ hopping.  The Hamiltonian is thus given by $H=\sum_\k h(\k)$ with:
\begin{align}
\label{Hamiltonian}
h(\k)=-t_0\left(
\begin{array}{cc}
h_{1}(\k) & h_{12}(\k)\\
 h_{12}^\dagger(\k)& h_{2}(\k) 
\end{array}
\right)\;.
\end{align}
where $h_{1/2}=E_0+\sigma_x+t/t_0[(c_1+c_2)\sigma_x+(s_1\mp s_2)\sigma_y]$  with $\sigma_x,\sigma_y$ the Pauli-matrices and $c_{1/2}=\cos(\k\cdot\a_{1/2})$, $s_{1/2}=\sin(\k\cdot\a_{1/2})$. We further set $E_0=0$. The coupling between the two dimers is given by
\begin{align}
h_{12}=\frac{t}{t_0}\left(
\begin{array}{cc}
1+e^{-i\k\cdot\a_1}& e^{-i\k\cdot(\a_1+\a_2)}+e^{-i\k\cdot\a_1}\\
1+e^{-i\k\cdot\a_2}& e^{-i\k\cdot(\a_1+\a_2)}+e^{-i\k\cdot\a_2}
\end{array}
\right)\;.
\end{align}
\begin{figure}[t]
\centering
  \includegraphics[width=1.05\columnwidth]{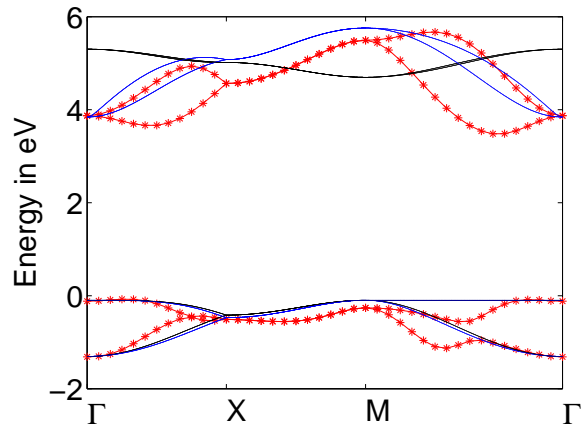}
\caption{(color online). Fit of the band structure obtained from the tight-binding model (black lines) for $t=t'=0.056t_0$ to the DFT-band structure (red stars). The blue lines resemble the extended tight-binding model with energy-dependent hopping terms.}
  \label{CompareDFT}
\end{figure}
In Fig. \ref{CompareDFT}, we compare the resulting band structure (black lines) to the full model obtained from DFT-calculations (red stars). Whereas the two valence bands (VBs) approximately agree with the exact first-principle band structure, the two conduction bands (CBs) do not show the large splitting in the $\Gamma-X$ and $\Gamma-M$ direction and seem to be inverted. Still, we want to emphasize that by setting $t_0=2.7$eV (which is the typical tight-binding hopping parameter of $\pi$-bands\cite{Neto09}) only one fitting parameter is involved by choosing $t/t_0=0.056$. Further, we applied a rigid energy shift of $E_0=2.3$eV. Also, the basic features of the band structure are reproduced, i.e., an (no) energy separation of the valence (conduction) bands at the $\Gamma$-point and a two-fold degeneracy along the $X-M$ direction. 

The conduction bands can be considerably improved within our four-band model by introducing an energy-dependent hopping parameter. This term is formally obtained by including an effective $p_z$-orbital of the sp$^3$-hybridized C1-atoms which are then projected out, see appendix. The projected 4x4 Hamiltonian is again given by Eq. (\ref{Hamiltonian}) with $t\rightarrow \frac{\tilde t_{C1-C2}^2}{E_X-E}$ where $E_X=E_{C1}^{pz}-E_{C2}^{pz}$ the effective energy difference of the $p_z$-atomic orbitals and $E_0=2\frac{\tilde t_{C1-C2}^2}{E_X-E}$. With $\tilde t_{C1-C2}=2$eV, $E_X=-3.75$eV and an additional rigid energy shift $E_0^c=1.15$eV, the eigenenergies of the conduction band are then obtained self-consistently to yield the upper blue lines of Fig. \ref{CompareDFT}. 
 
In order to justify the above parameters, it is necessary to include an onsite interaction at the C2-atoms and assisted hopping terms, see appendix. Neglecting these correlations, the onsite energy shift yields $E_X=2$eV and the two valence bands of the original four-band model with constant hopping amplitude $t=0.056t_0$ are almost perfectly reproduced by the four-band model with energy-dependent hopping parameter of $\tilde t_{C1-C2}=0.35t_0$. Additionally, we introduced an energy shift of $E_0^v=2.61$eV. This can be seen in Fig. \ref{CompareDFT} where the lower blue lines are almost on top of the black lines. 

Our analysis indicates that correlation effects are important in order to reproduce the band-structure of the conduction band obtained from DFT-calculations and that electron-electron interactions are effectively screened in the valence band. The change in $\tilde t_{C1-C2}$ and the different constant energy shifts $E_0^c$ and $E_0^v$ further suggest an assisted hopping amplitude which depends on the occupation number of the C2-atoms.\cite{Hirsch91,Hirsch93} 

The inclusion of both correlation terms yields a consistent picture when choosing the Hubbard-interaction $U\approx10$eV and the assisted hopping term $W\approx2$eV. The assisted hopping term may further lead to a superconducting condensate when pumping electrons into the conduction bands.\cite{Hirsch91,Hirsch93,Guinea03,Stauber04}  

\subsection{Analytical solution}
The eigenenergies of the bands around the neutrality point can be well approximated by the (extended) 4x4 model, reproducing correctly the degeneracy of the VBs and CBs along the $X-M$-direction and the band splitting along the other directions with a (zero) gap in the VBs (CBs). Nevertheless, to also obtain orthonormal eigenvectors, we will continue the discussion by approximating the energy-dependent hopping parameter by the constant hopping parameter $t$ also for the conduction bands. 

In the following, we set $t_0=1$ and assume $t\ll1$. A first approximation is thus given by neglecting the inter-dimer coupling $h_{12}$. Comparing this approximation with the exact numerical solution, we see that the conduction bands are well described by neglecting $h_{12}$. For the eigenvalues, we then obtain $\epsilon_\pm^c=|z_\pm|$ with
\begin{align}
z_\pm=1+(c_1+c_2)t+i(s_1\pm s_2)t\;.
\end{align}
However, the valence band experiences a splitting that cannot be account for by simply setting $h_{12}=0$. Let thus $U$ denote the unitary transformation which diagonalizes the Hamiltonian with $h_{12}=0$ and lets only consider the upper left and lower right $2\times2$-matrix  of $U^\dagger HU$ that connects the two valence and conduction band states, respectively. Keeping now only terms of these matrices that are linear in $t$ yields the linear approximation for $\epsilon_\pm^c$ and the following eigenvalues for the valence band:

\begin{align}
\epsilon_\pm^v&=-1-(c_1+c_2)t\pm|z_0| t\;,\;\text{with}\\
z_0&=(1+c_1-is_1)(1+c_2-is_2)\notag
\end{align}
The corresponding eigenvectors need to be transformed by $U$ to obtain the approximate eigenstates of the original Hamiltonian. We thus end up with the following set of 4-dimensional eigenvectors:
\begin{align}
\label{eigenvectors}
|\k,-\rangle_c&= \frac{1}{\sqrt{2}}\left( \begin{array}{c}
-e^{-i\varphi_-}\\
1\\
0\\
0
\end{array} \right),\notag\\
|\k,+\rangle_c&= \frac{1}{\sqrt{2}}\left( \begin{array}{c}
0\\
0\\
-e^{-i\varphi_+}\\
1
\end{array} \right),\notag\\
|\k,\mp\rangle_v&= \frac{1}{2}\left( \begin{array}{c}
\pm e^{i\vartheta}e^{-i\varphi_-}\\
\pm e^{i\vartheta}\\
 e^{-i\varphi_+}\\ 
1
\end{array} \right)\;.
\end{align}
with $e^{i\varphi_\pm}=\frac{z_\pm}{|z_\pm|}$ and $e^{i\vartheta}=\frac{z_0}{|z_0|}$. 

\begin{figure}[t]
\centering
  \includegraphics[width=1.05\columnwidth]{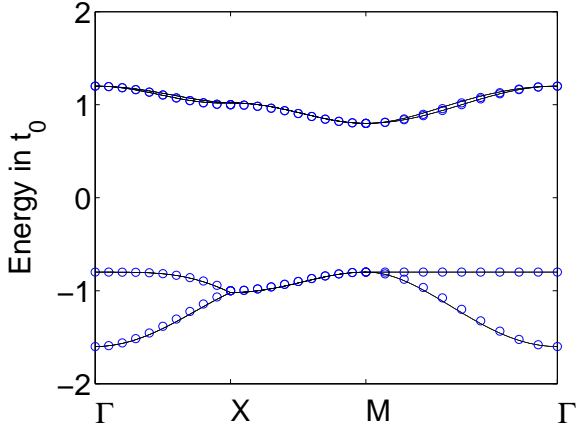}
\caption{(color online): Band structure of the four-band tight-binding model with constant hopping amplitude $t=t'=0.1t_0$ and $E_0=0$ (black line) compared to the analytical approximation (blue circles).}
  \label{CompareAnalytic}
\end{figure}

The above approximation matches excellently the numerically exact band structure for $t/t_0\ll1$, see Fig. \ref{CompareAnalytic}.

\section{Optical absorption}
Penta-graphene displays a band gap of 3.25eV which makes it a potential candidate for blue absorption/emission. In order to discuss the optical response of the system, we will obtain the coupling Hamiltonian via the Peierls substitution in $\k$-space. Nevertheless, for systems with various atoms in the unit cell, it is crucial to represent the Hamiltonian within a basis that distinguishes the relative phase of the atoms within the same unit cell.\cite{Paul03}
The above representation has thus to be modified by using the $\k$-states of the following basis:
\begin{widetext}
\begin{align}
|\k\rangle=\sum_{m,n} e^{i\k\cdot(m\a_1+n\a_2)}\left[\alpha(\k)e^{i\k\cdot\delta_A}a_{mn}^\dagger+\beta(\k)e^{i\k\cdot\delta_B}b_{mn}^\dagger+\gamma(\k)e^{i\k\cdot\delta_C}c_{mn}^\dagger+\delta(\k)e^{i\k\cdot\delta_D}d_{mn}^\dagger\right]
\end{align}
\end{widetext}
Placing the origin of the unit cell in the middle of the AB-dimer and choosing it in the direction of the $y$-axis, we have $\delta_A=(0,-b/2)$, $\delta_B=(0,b/2)$, $\delta_C=(b_+,0)$, and $\delta_D=(b_++b,0)$ where $b_+=\frac{a}{2}(\sqrt{3}+1)$. For the regular Cairo patterning, we have $b=a(\sqrt{3}-1)$; the unit-cell vectors are $\a_1=\tilde a(1,1)$ and $\a_2=\tilde a(1,-1)$ with $\tilde a=\sqrt{3}a$. 

With respect to the new basis, the 2x2 Hamiltonians of Eq. (\ref{Hamiltonian}) now read
\begin{widetext}
\begin{align}
h_1&=\left(
\begin{array}{cccc}
0&e^{ik_yb}+2\frac{t}{t_0}e^{-ik_ya}\cos(k_x\tilde a)\\
e^{-ik_yb}+2\frac{t}{t_0}e^{ik_ya}\cos(k_x\tilde a)&0
\end{array}\right)\;,
\\
h_2&=\left(
\begin{array}{cccc}
0&t_0e^{ik_xb}+2\frac{t}{t_0}e^{-ik_xa}\cos(k_y\tilde a)\\
t_0e^{-ik_xb}+2\frac{t}{t_0}e^{ik_xa}\cos(k_y\tilde a)&0
\end{array}\right)\;,
\\
h_{12}&=\frac{t}{t_0}\left(
\begin{array}{cc}
  e^{i\k\cdot(b_+,b_-)}+e^{-i\k\cdot(b_-,b_+)}&e^{-i\k\cdot(b_+,-b_-)}+e^{i\k\cdot(b_-,-b_+)}\\
 e^{i\k\cdot(b_+,-b_-)}+e^{-i\k\cdot(b_-,-b_+)}&
e^{-i\k\cdot(b_+,b_-)}+e^{i\k\cdot(b_-,b_+)} 
\end{array}\right)\;.
\end{align}
\end{widetext}
The eigenvectors read
\begin{align}
\label{eigenvectors}
|\k,-\rangle_c&= \frac{1}{\sqrt{2}}\left( \begin{array}{c}
-e^{-i\varphi_-}e^{-i\k\cdot\delta_A}\\
e^{-i\k\cdot\delta_B}\\
0\\
0
\end{array} \right),\\
|\k,+\rangle_c&= \frac{1}{\sqrt{2}}\left( \begin{array}{c}
0\\
0\\
-e^{-i\varphi_+}e^{-i\k\cdot\delta_C}\\
e^{-i\k\cdot\delta_D}
\end{array} \right),\notag\\
|\k,\mp\rangle_v&= \frac{1}{2}\left( \begin{array}{c}
\pm e^{i\vartheta}e^{-i\varphi_-}e^{-i\k\cdot\delta_A}\\
\pm e^{i\vartheta}e^{-i\k\cdot\delta_B}\\
 e^{-i\varphi_+}e^{-i\k\cdot\delta_C}\\ 
e^{-i\k\cdot\delta_D}
\end{array} \right)\;.
\end{align}

\subsection{Fermi's "golden rule"}
We are interested in the system response at small fields and therefore expand the Hamiltonian up to terms linear in $\k$. With $\k\rightarrow\k+\frac{e}{\hbar}\bf A$ (minimal coupling), we obtain the following coupling Hamiltonian:
\begin{align}
\label{Coupling}
V=-\left(
\begin{array}{cc}
v_{1} & v_{12}\\
 v_{12}^\dagger& v_{2} 
\end{array}
\right)\;,
\end{align}
with $v_1=ev\sigma_yA_y(1-2\frac{t}{t_0}\frac{a}{b})$, $v_2=ev\sigma_yA_x(1-2\frac{t}{t_0}\frac{a}{b})$ and
\begin{align}
v_{12}=ie\tilde v\left(
\begin{array}{cc}
A_x-A_y&-A_x-A_y\\
A_x+A_y&-A_x+A_y
\end{array}\right)
\end{align}
where the velocities $\hbar v=bt_0$ and $\hbar\tilde v=at$.

We can now apply Fermi's "golden rule" to calculate the absorption due to an incoming electric field $\E(t)=-\partial_t\bf A$. The incoming energy flux of a propagating sinusoidal linearly polarized electromagnetic plane wave of a fixed frequency is given by $W_i=\frac{\epsilon_0c}{2}|E_0|^2$ and the absorbed energy flux $W_a=\eta\hbar\omega$ with $\eta$ the transition rate. Since the momentum is conserved in the absorption process, only transitions from $(\k,v)$ to $(\k,c)$ are allowed. The total transition rate is then obtained by summing over all initial states $\k$ which yields:
\begin{align}
\eta&=\frac{2\pi}{4\hbar}\frac{g_s}{A}\sum_{\k;m,n=\pm}\left|{}_c\langle \k,n|V|\k,m\rangle_v\right|^2\delta(\hbar\omega-\epsilon_n^c+\epsilon_m^v)
\end{align}
where the sum goes over the first Brillouin zone. For a  system with $N_c=N^2$ unit cells, the possible $\k$-values thus read:
\begin{align}
\k=\frac{n_1}{N}\b_1+\frac{n_2}{N}\b_2
\end{align}
with $\b_{1/2}=\frac{\pi}{3a}(\sqrt{3},\pm\sqrt{3})$ and $n_{1/2}=0,...,N-1$.

Only considering the leading term in $t_0$, we can neglect $v_{12}$ and obtain for the matrix-elements:
\begin{align}
\left|{}_c\langle \k,-|V|\k,m\rangle_v\right|^2=\left(\frac{evE_y}{\sqrt{2}\omega}\right)^2\cos^2(\varphi_--k_yb)\;,\\
\left|{}_c\langle \k,+|V|\k,m\rangle_v\right|^2=\left(\frac{evE_x}{\sqrt{2}\omega}\right)^2\cos^2(\varphi_+-k_xb)\;.
\end{align}

By interchanging $k_x\leftrightarrow k_y$, we have $\varphi_+\leftrightarrow\varphi_-$, indicating that the absorption displays the underlying four-fold symmetry within our approximation (degenerate conduction band). But we will neglect $\varphi_\pm$ in the following since it would yield another correction of order $t$. We can thus write the transition rate as
\begin{align}
\label{eta}
\eta&=\frac{2\pi}{8\hbar}\left(\frac{ev}{\omega}\right)^2\frac{g_s}{A}\sum_{\k,m=\pm}\delta(\hbar\omega-\epsilon^c+\epsilon_m^v)\\\notag&\times\left(E_x^2\cos^2(k_xb)+E_y^2\cos^2(k_yb)\right)\;.
\end{align}

\subsection{Optical absorption}
One measure for absorption is the joint density of states (DOS):
\begin{align}
\label{jointDOS}
	\rho^{cv}(\omega)=\frac{2g_s}{A}\sum_{\k,m}\delta(\hbar\omega-\epsilon^c+\epsilon_m^v)\;, 
\end{align}
where the 2 comes from the two-fold degeneracy of the conduction band.

The DOS of the (degenerate) conduction band $\rho^c$ is also given by Eq. (\ref{jointDOS}) by setting $\epsilon_m^v=0$ and neglecting the summation over $m$. With $\hbar\tilde\omega=(\hbar\omega-t_0)/(2t)$, we can obtain an analytical expression:
\begin{align}
	\rho^c(\omega)=\frac{2g_s}{3\pi^2a^2t}K(1-(\hbar\tilde\omega)^2)\Theta(1-(\hbar\tilde\omega)^2)\;,
\end{align}
where $K(x)$ is the complete elliptic function. At the band-edges, we thus have a constant density of state of $\rho^c=\frac{g_s}{3\pi a^2t}$. At the band-center, there is a van-Hove singularity at the $X$-point with a logarithmically diverging density of states $\rho^c\to\frac{g_s}{3\pi^2 a^2t}\ln\hbar\tilde\omega$.

The DOS of the valence band (setting $\epsilon^c=0$) can also be expressed in closed form, but already this and especially the joint density of states are quite complicated analytical function and we will not pursue this any further. Instead, we will expand the dispersion around the high-symmetry points and calculate the optical absorption which should be particularly high at van-Hove singularities. 

The optical absorption is defined by the ratio of the absorbed and the incoming energy flux, $\mathcal{A}=\frac{W_a}{W_i}$. The expansion around the $\Gamma$-point ($\Gamma=0$) yields an isotropic absorption for both transitions from the two $(m=\pm)$-valence bands. These are related to the transition energies $\hbar\omega_+=2t_0$ and $\hbar\omega_-=2t_0+8t$, respectively. To leading order in $t_0/t$, we get:
\begin{align}
\mathcal{A}_+&=\pi\alpha\frac{g_s}{9}\left(\frac{b}{a}\right)^2\frac{t_0}{t}\approx3.5\mathcal{A}_0\\
\mathcal{A}_-&=\pi\alpha\frac{g_s}{3}\left(\frac{b}{a}\right)^2\frac{t_0}{t}\approx10.5\mathcal{A}_0
\end{align}
The numerical values are obtained from $(b/a)^2\approx0.88$ and $t_0/t\approx18$ in units of the universal absorption of suspended graphene $\mathcal{A}_0=\pi\alpha\approx2.3\%$ with $\alpha=\frac{e^2}{4\pi\epsilon_0\hbar c}$ the fine-structure constant.\cite{Nair08,Mak08} This universal absorption is also present in InAs-quantum wells\cite{Fang13} and other systems\cite{Stauber15}. But here, the expression depends on material constants and is thus non-universal. It is also substantially higher with $\mathcal{A}_-\approx24\%$. This is remarkable having in mind that the maximal absorption of suspended two-dimensional materials is 50 \% and to our knowledge the highest value of a suspended two-dimensional system.\cite{deAbajo07} However, at such high absorption the simple "golden rule"-approach cannot be trusted anymore and more accurate calculations involving the optical conductivity are needed.

We finally note that the absorption will in general depend on the polarisation of the incident light, as can be appreciated from Eq. (\ref{eta}). For transitions from the $m=-1$-valence band to the conduction band around the $M$-point, we obtain, e.g., in leading order in $t_0/t$
\begin{align}
\mathcal{A}_-^M=\pi\alpha\frac{g_s}{9}\left(\frac{b}{a}\right)^2\frac{t_0}{t}\left[1-\tilde s^2\cos^2\beta\right]\;,
\end{align}
with $\tilde s=\sin(\frac{\pi}{\sqrt{3}}\frac{b}{a})\approx1$ and $\beta$ the angle of the polarisation vector with the $x$-axis. There is thus almost no absorption for $y$-polarized incident light yielding an asymmetry ratio of nearly 100\%. Moreover, ferromagnetism is expected due to the nearly flat band.\cite{Mielke93}

\section{Summary}
In this work, we have investigated the possibility of an analytical description of penta-graphene. We were able to adequately approximate the two highest valence and lowest conduction bands within a simple four band model, but to match the conduction bands an energy dependent hopping element was necessary. To justify the parameters of the effective model, the inclusion of an onsite energy at the sp$^2$-hybridized C2-atoms was necessary. This interaction should be screened for the valence band and we also expect an assisted hopping term in the effective model which might lead to a superconducting condensate at low temperatures when pumping electrons into the conduction bands.\cite{Hirsch91,Hirsch93,Guinea03,Stauber04} A possible extension of our approach would include multi-orbitals in the spirit of the Weaire-€"Thorpe (WT) model.\cite{Weaire71,Hatsugai15}

Using a constant hopping element between dimers, we were able to present an analytical solution for the eigenvalues as well as for the eigenvectors and calculated the optical absorption within this approximation. For small field strengths and for transitions around the $\Gamma$-point, this yielded an isotropic absorption of up to 24\% which is remarkably large compared to usual two-dimensional materials such as graphene.\cite{Stauber15} For transitions away from the $\Gamma$-point, we expect an absorption which strongly depends on the polarisation of the incident light. 

It would be interesting to further investigate the influence of correlation effects on the band-structure as well as on the optical properties. Also the inclusion of electron-hole interaction and excitonic resonances are likely to change the absorption close to the band edges. These issues shall be discussed in the future.

\begin{acknowledgments} 
We thank F\'elix Yndurain and Paul Wenk for useful discussions. This work has been supported by Spain's MINECO under grant FIS2013-48048-P, and by Deutsche
Forschungsgemeinschaft via GRK 1570. JIB thanks the ERC starting Investigator Award, grant \#239739 STEMOX.
\end{acknowledgments} 

\appendix
\section{Effective Hamiltonian including correlation effects}
In order to  adequately describe the conduction band, the C1-atoms as well as correlation effects need to be included. The minimal model consists of one (effective) $p_z$-orbital for each carbon atom. Additionally, a Hubbard interaction at the C2-atoms as well as an assisted hopping term is included where the hopping amplitude depends on the occupation number of the C2-atoms.\cite{Hirsch91,Hirsch93} On the other hand, we neglect the onsite interaction at the C1-atoms because the effective $p_z$-orbital is spread out due to the sp$^3$-hypridization. The effective model thus reads
\begin{align}
\mathcal{H}=\mathcal{H}_{0}+\mathcal{H}_{C}+\mathcal{H}_{H}+\mathcal{H}_{a}
\end{align}
where we have defined the bare hopping Hamiltonian
\begin{align}
&\mathcal{H}_0=-t_0\sum_{\mathclap{m,n;\sigma}}\left(a^\dagger_{m,n;\sigma}b_{m,n;\sigma}+c^\dagger_{m,n;\sigma}d_{m,n;\sigma}+H.c.\right)\notag\\
&-t_{C1-C2}\sum_{\mathclap{m,n;\sigma}}\left(a^\dagger_{m,n;\sigma}(e_{m,n;\sigma}+f_{m,n-1;\sigma})+H.c.\right)\notag\\
&-t_{C1-C2}\sum_{\mathclap{m,n;\sigma}}\left(b^\dagger_{m,n;\sigma}(e_{m-1,n;\sigma}+f_{m,n;\sigma})+H.c.\right)\notag\\
&-t_{C1-C2}\sum_{\mathclap{m,n;\sigma}}\left(c^\dagger_{m,n;\sigma}(e_{m,n;\sigma}+f_{m,n;\sigma})+H.c.\right)\notag\\
&-t_{C1-C2}\sum_{\mathclap{m,n;\sigma}}\left(d^\dagger_{m,n;\sigma}(e_{m,n+1;\sigma}+f_{m+1,n;\sigma})+H.c.\right)\label{hoppingH}\;,
\end{align}
where $e$ and $f$ denote the lower and upper red atom in the unit cell of Fig. \ref{Cairo}a). The four sp$^2$-hybridized atoms are defined in Fig.  \ref{Cairo}b)

The onsite Hamiltonian given by 
\begin{align}
\mathcal{H}_{C}&=\sum_{\mathclap{\substack{m,n;\sigma\\g=a,b,c,d,e,f}}}E_gn^g_{m,n;\sigma}\;,
\end{align}
with $E_g=E_{C1}$ for $g=e,f$ and $E_g=E_{C2}$ for $g=a,b,c,d$ and $n^g_{m,n;\sigma}=g^\dagger_{m,n;\sigma}g_{m,n;\sigma}$ for $g=a,b,c,d,e,f$.

We also need to introduce a Hubbard term
\begin{align}
\mathcal{H}_{H}=U\sum_{\mathclap{\substack{m,n\\g=a,b,c,d}}}n^g_{m,n;\uparrow}n^g_{m,n;\downarrow} 
\end{align}
and an assisted hopping contribution 
\begin{align}
&\mathcal{H}_a=W\sum_{\mathclap{m,n;\sigma}}n^a_{m,n;\bar\sigma}\left(a^\dagger_{m,n;\sigma}(e_{m,n;\sigma}+f_{m,n-1;\sigma})+H.c.\right)\notag\\
&+W\sum_{\mathclap{m,n;\sigma}}n^b_{m,n;\bar\sigma}\left(b^\dagger_{m,n;\sigma}(e_{m-1,n;\sigma}+f_{m,n;\sigma})+H.c.\right)\notag\\
&+W\sum_{\mathclap{m,n;\sigma}}n^c_{m,n;\bar\sigma}\left(c^\dagger_{m,n;\sigma}(e_{m,n;\sigma}+f_{m,n;\sigma})+H.c.\right)\notag\\
&+W\sum_{\mathclap{m,n;\sigma}}n^d_{m,n;\bar\sigma}\left(d^\dagger_{m,n;\sigma}(e_{m,n+1;\sigma}+f_{m+1,n;\sigma})+H.c.\right)\label{AssistedHopping}\;,
\end{align}
where $\bar\sigma$ is the opposite spin-projection of $\sigma$.
 
The interaction terms are most easily treated within the mean-field approximation. For the Hubbard interaction, we set
\begin{align}
\label{MeanField}
\mathcal{H}_{H}\approx
U\sum_{\mathclap{\substack{m,n;\sigma\\g=a,b,c,d}}}n^g_{m,n;\sigma}\langle n^g_{m,n;\bar\sigma}\rangle+E_U\;,
\end{align}
where the constant energy shift reads $E_U=-U\sum\langle n^g_{m,n;\uparrow}\rangle \langle n^g_{m,n;\downarrow}\rangle$ with the sum over $m,n$ and $g=a,b,c,d$. The assisted hopping term is approximated analogously by the following:
\begin{widetext}
\begin{align}
\mathcal{H}_a&=W\sum_{\mathclap{m,n;\sigma}}\langle n^a_{m,n;\bar\sigma}\rangle\left(a^\dagger_{m,n;\sigma}(e_{m,n;\sigma}+f_{m,n-1;\sigma})+H.c.\right)
+\langle n^b_{m,n;\bar\sigma}\rangle\left(b^\dagger_{m,n;\sigma}(e_{m-1,n;\sigma}+f_{m,n;\sigma})+H.c.\right)\notag\\
&+W\sum_{\mathclap{m,n;\sigma}}\langle n^c_{m,n;\bar\sigma}\rangle\left(c^\dagger_{m,n;\sigma}(e_{m,n;\sigma}+f_{m,n;\sigma})+H.c.\right)
+\langle n^d_{m,n;\bar\sigma}\rangle\left(d^\dagger_{m,n;\sigma}(e_{m,n+1;\sigma}+f_{m+1,n;\sigma})+H.c.\right)\notag\\
&+W\sum_{\mathclap{m,n;\sigma}}n^a_{m,n;\bar\sigma}\left\langle a^\dagger_{m,n;\sigma}(e_{m,n;\sigma}+f_{m,n-1;\sigma})+H.c.\right\rangle+n^b_{m,n;\bar\sigma}\left\langle b^\dagger_{m,n;\sigma}(e_{m-1,n;\sigma}+f_{m,n;\sigma})+H.c.\right\rangle\notag\\
&+W\sum_{\mathclap{m,n;\sigma}}n^c_{m,n;\bar\sigma}\left\langle c^\dagger_{m,n;\sigma}(e_{m,n;\sigma}+f_{m,n;\sigma})+H.c.\right\rangle+n^d_{m,n;\bar\sigma}\left\langle d^\dagger_{m,n;\sigma}(e_{m,n+1;\sigma}+f_{m+1,n;\sigma})+H.c.\right\rangle+E_W,
\end{align}
\end{widetext}
where the constant energy shift reads $E_W=-W\sum\langle n^g_{m,n;\bar\sigma}\rangle \langle \xi\rangle$ with the sum over $m,n;\sigma;g=a,b,c,d$ and $\xi=C1^\dagger C2+H.c.$ denoting all the different hopping processes between the C1- and C2-atoms which are proportional to the hopping amplitude $t_{C1-C2}$. 

For a half-filled band, we have $\langle n^g_{m,n;\sigma}\rangle=1/2$ leading to the following approximation:
\begin{align}
\mathcal{H}_{C}+\mathcal{H}_{H}+&\mathcal{H}_a\approx\\\notag
E_X\sum_{\mathclap{m,n;\sigma}}\left(n^e_{m,n;\sigma}+n^f_{m,n;\sigma}\right)& 
+\mathcal{\widetilde H}_0+\sum_{\mathclap{m,n;\sigma}} \widetilde E_W\;,
\end{align}
where we have $E_X=E_{C1}-E_{C2}-U/2-W\langle \xi\rangle/8$ and $\widetilde E_W=-W\langle \xi\rangle/4$. We have also set $E_{C2}=0$. The assisted hopping Hamiltonian further leads to a renormalized hopping amplitude $\tilde t_{C1-C2}= t_{C1-C2}+W/2$ and $\mathcal{\widetilde H}_0$ is thus obtained by replacing $t_{C1-C2}$ by $\tilde t_{C1-C2}$ in Eq. (\ref{hoppingH}). 

The expectation value of the hopping processes is obtained self-consistently as $\langle \xi\rangle=2.94$. For typical parameters, we have $U\approx10$eV, $W\approx2$eV, $E_{C1}\approx2$eV and set $E_{C2}=0$ which yields $E_X\approx -3.75$eV and $\widetilde E_W\approx-1.5$eV. We further set $t_{C1-C2}=0.35t_0\approx1$eV which yields $\tilde t_{C1-C2}\approx2$eV.

Within this mean-field approximation, the two spin-projections decouple and we will drop the spin-degree of freedom. We are now in the position to reduce the resulting 6-band model to an effective 4-band model by projecting out the C1-atoms. This results in the effective Hamiltonian of  Eq. (\ref{Hamiltonian}) with $t\rightarrow \frac{\tilde t_{C1-C2}^2}{E_X-E}$ and $E_0=2\frac{\tilde t_{C1-C2}^2}{E_X-E}$. With an additional shift of $E_0^c=1.15$eV, the two conduction bands can be well approximated with the above parameters. 

The electronic density of the two valence bands is more located between the C2-atoms than the one of the two conductance band. Also self-energy corrections are usually more dominant for the bands further away of the Fermi level, i.e., in our case the conduction band. The two valence bands are thus calculated by setting $U=W=0$. The larger energy shift of $E_0^v=2.61$eV needed to fit the data matches well with the predicted value $E^c-E^v\approx \widetilde E_W$. The four bands are shown as  blue curves in Fig. \ref{CompareDFT}. 

We note that the hopping Hamiltonian $\mathcal{H}_0$ has already been discussed in the Supplementary Information S3 of Ref. \onlinecite{Zhang15}. Here, we showed that it is crucial to also include the onsite, Hubbard and assisted hopping terms in order to reproduce the band structure of the bands close to half-filling.

\end{document}